\documentstyle[aaspp4]{article}

\newcommand{\kms}{km~s$^{-1}$}
\newcommand{\hii}{\ion{H}{2}}
\newcommand{\ha}{H$\alpha$}
\newcommand{\vlsr}{v_{\rm LSR}}
\newcommand{\dist}{d_1}
  
\begin{document}

\title{Faint, Large-scale \ha\ Filaments in the Milky Way}

\author{L. M. Haffner\\\texttt{haffner@astro.wisc.edu}}
\author{R. J. Reynolds\\\texttt{reynolds@astro.wisc.edu}}
\author{S. L. Tufte\\\texttt{tufte@astro.wisc.edu}}
\affil{Department of Astronomy, University of Wisconsin--Madison\\475
  North Charter Street, Madison, WI 53706}

\begin{abstract}
  During the initial data reduction of the Wisconsin H-Alpha Mapper
  (WHAM) \ha\ Sky Survey,
  we have discovered several very long ($\sim 30\arcdeg$--80\arcdeg)
  filaments superimposed on the diffuse \ha\ background. These
  features have no clear correspondence to the other phases of the
  interstellar medium revealed by 21 cm, X-ray, IR, or radio continuum
  surveys, and they have no readily identifiable origin or source of
  ionization. In this letter, the data for two of these faint
  ($I_{H\alpha} \approx$ 0.5--1.5 R) structures are presented. The
  first is an 80\arcdeg-long, 2\arcdeg-wide arch that extends nearly
  perpendicular to the Galactic plane at $\ell = 225\arcdeg$ and
  attains a maximum latitude of $+51\arcdeg$ near $\ell = 240\arcdeg$
  before reaching the southern boundary of our survey map at $\ell =
  270\arcdeg$, $b = +42\arcdeg$.  The vertical portion of this feature
  between $b = +10\arcdeg$ and $+25\arcdeg$ is associated with a
  single radial velocity component centered at $\vlsr = +16$ \kms\ 
  with a full width at half maximum of 27 \kms. A decrease in the
  velocity is observed from $b = +33\arcdeg$ through $+48\arcdeg$ as
  the feature arches toward higher Galactic longitudes. At this end,
  the emission component is centered near $\vlsr = -20$ \kms.  Where
  this feature appears to meet the Galactic plane near $\ell =
  225\arcdeg$, it is directly above the \hii\ region surrounding CMa
  R1/OB1.  A second filament consists of a $\sim
  25\arcdeg$--30\arcdeg-long arc spanning $\ell =
  210\arcdeg$--240\arcdeg\ at $b = +30\arcdeg$ to 40\arcdeg.  The
  radial velocity of this feature increases systematically from 0
  \kms\ at $\ell = 215\arcdeg$, $b = +38\arcdeg$ to +18 \kms\ at $\ell
  = 236\arcdeg$, $b = +28\arcdeg$. Both features have rather constant
  intensities along their entire lengths, ranging from 0.5--1.5 R (EM
  = 1--3 cm$^{-6}$ pc) with no obvious trends.
\end{abstract}


\section{Introduction}
\label{sec:intro}

The Wisconsin H-Alpha
Mapper\footnote{\texttt{http://www.astro.wisc.edu/wham/}} (WHAM)
survey is providing the first velocity-resolved map of the \ha\ 
emission from our Galaxy's diffuse interstellar medium.  The
combination of WHAM's sensitivity ($< 0.1$ R) and velocity resolution
(12 \kms) reveals new details about the large-scale structure and
kinematics of the Warm Ionized Medium (WIM).  Studies of faint
emission structures in the WIM may help us understand the distribution
and morphology of the ionized gas and may even provide direct evidence
of ionization sources. In this letter, we present the discovery of two
large, faint filaments in our Galaxy.

\section{Observations}
\label{sec:obs}

WHAM consists of a dedicated 0.6-m, all-sky siderostat connected to a
15-cm, double-etalon Fabry-Perot spectrometer.  The optical design
delivers a spectrum covering a 200 \kms\ radial velocity interval from
a one-degree circular patch of sky. For the \ha\ sky survey,
approximately 35,000 spectra were obtained above $\delta = -20\arcdeg$
with the one-degree diameter beam centered on a $0\fdg98 \times
0\fdg85$ grid in $\ell$ and $b$. The integration time for each
one-degree pixel was 30 seconds, resulting in a signal-to-noise ratio
of about 30 in background continuum regions of the spectrum.  This
allows the detection of sources as faint as 0.1 R (1 R $= 10^6/4\pi $
ph cm$^{-2}$ s$^{-1}$ ster$^{-1}$ or $2.4 \times 10^{-7}$ ergs
cm$^{-2}$ s$^{-1}$ ster$^{-1}$ at \ha) in a single exposure.  The
etalons are configured to provide a spectral resolution of 12 \kms,
which is sufficient to identify and remove atmospheric lines in the
spectra and fully resolve the \ha\ emission-line profiles from warm
gas ($T \sim 8000$ K) in the Galaxy.  A detailed description of the
instrument design is given by Tufte (1997).

Presented here are nearly 3800 of these spectra extracted from the
WHAM \ha\ Sky Survey. They cover the region of the sky from $\ell =
200\arcdeg$ to $270\arcdeg$ and $b = -10\arcdeg$ to $+60\arcdeg$ above
$\delta = -20\arcdeg$. Most of these spectra were observed by WHAM
from 1997 February through 1997 May. These observations are designated
``O1'' below. A small portion (145 spectra) from $\ell = 253\arcdeg$
to $270\arcdeg$ and $b = +47\arcdeg$ to $+60\arcdeg$ was observed in
January 1998 and are labeled ``O2'' below. Standard WHAM data
reduction, including bias subtraction, reflected ring subtraction,
annular-summing, and flat fielding, was used to create these spectra
(\cite {hrt98}).

Atmospheric emission lines are a significant source of contamination
in WHAM spectra. The brightest of these, with intensities ranging from
2--13 R, is the geocoronal \ha\ line emitted in the earth's upper
atmosphere. Since the earth's orbital velocity changes the location of
the velocity frame of the Local Standard of Rest (LSR) with respect to
the geocentric velocity frame, most directions in the sky were
observed when the geocoronal line was separated by at least 20 \kms\ 
from the LSR. This separation makes it easy to resolve the relatively
narrow (6--8 \kms\ FWHM) geocoronal line from the Galactic \ha\ 
emission and minimize its effect. The data presented here have
velocity separations of the geocoronal line from the LSR that range
from $-24$ \kms\ to $-42$ \kms\ (O1) and $+14$ \kms\ to $+25$ \kms\ 
(O2). A single Gaussian component has been fitted and subtracted from
these spectra to remove the geocoronal line.

Two other much fainter atmospheric emission lines, believed to be
extremely weak OH emission, are present in the spectra. The first is
located 71 \kms\ to the red of the geocoronal line ($\vlsr=+27$ to
$+45$ \kms\ in O1); it has a width of 23.5 \kms\ and varies in
intensity from 0.05 to 0.36 R. Deeper observations and the wide line
width suggest that this feature may be a pair of atmospheric lines.
The second is located 38 \kms\ to the blue of the geocoronal line
($\vlsr=-64$ to $-82$ \kms\ in O1); it has a width of 9.4 \kms\ and
varies in intensity from 0.04 to 0.13 R. For this study, the red sky
line provides the most contamination near the LSR in O1, and it has
been removed by approximating it as a single Gaussian component.
Variations in the line intensities are most severe from night to
night.  Our corrections account for intensity variations over
30-minute periods, which are typically 0.02 R. Other faint atmospheric
lines are present in O2 even further to the blue from the geocoronal
line. These features will not be discussed here since they do not
affect our study near the LSR.

A first-order polynomial fits the sky background well within our 200
\kms\ bandpass. The typical level of the subtracted background is
approximately 0.03--0.07 R (\kms)$^{-1}$ or 1.4--3.2 R \AA$^{-1}$,
which is dominated by atmospheric continuum emission above $|b| =
30\arcdeg$.  Individual stars within the one-degree beam brighter than
sixth magnitude contribute significantly to the background in several
directions. Some of the spectra in these directions also contain an
\ha\ absorption line from the stars. These pointings are most
noticeable in Figure~\ref{fig:images} as isolated one-degree diameter
depressions in the otherwise smooth emission at higher Galactic
latitudes.

Absolute intensity calibration is accomplished by frequent
observations of nebular sources. All WHAM intensities are tied to the
absolute intensity measurement of the North American Nebula (NAN; NGC
7000), 850 R within a 50$\arcmin$ beam centered at $\ell = 85\fdg60$,
$b = -0\fdg71$, as determined by Scherb (1981). Since we have yet to
apply the gamut of intensity calibrations to our data, relative
intensities quoted here are accurate to about 10\%. The systematic
uncertainty in the absolute scale (i.e., the intensity of NGC 7000) is
believed to be approximately 15\%.

\section{Results}
\label{sec:results}

Figure~\ref{fig:images} displays four velocity integrated images of
\ha\ emission from our selected region of the sky. The two filaments
of interest are most easily seen in the narrow (12 \kms) velocity
interval images, although their overall structure can also be seen on
the ``broad-band'' map (Figure~\ref{fig:images}a). The vertical
portion of the longest filament (hereafter Filament 1) is most
prominent in the map centered on $\vlsr = +20$ \kms\ 
(Figure~\ref{fig:images}d), stretching nearly perpendicular to the
Galactic plane along $\ell = 225\arcdeg$. It then extends toward
higher Galactic longitudes, turning over near $b = +50\arcdeg$ while
shifting to more negative velocities in Figures \ref{fig:images}c and
\ref{fig:images}b. A second long filament (Filament 2) can been seen
as an arc extending across the images near $b = +30\arcdeg$ to
$+40\arcdeg$.  Its peak emission moves from $\ell = 220\arcdeg$ in the
$\vlsr = -20$ \kms\ map (Figure~\ref{fig:images}b) to $\ell =
235\arcdeg$ in the $\vlsr = +20$ \kms\ map (Figure~\ref{fig:images}d).
Other filaments are also apparent, including the feature running from
$\ell = 205\arcdeg$, $b = +25\arcdeg$ to $\ell = 220\arcdeg$, $b =
+30\arcdeg$. In this letter we focus primarily on the 80\arcdeg-long
Filament 1.

An $\ell$-$v$ map is presented in Figure~\ref{fig:lv/bcut} along $b =
+25\arcdeg$. From this representation, Filament 1 appears to be a
narrow ($\lesssim 2\arcdeg$) feature at $\ell = 225\arcdeg$ that is
offset in velocity from the smooth emission present near the LSR at
other Galactic longitudes. An analysis of the spectra reveal that this
velocity offset is about $+18$ \kms. The spectrum of Filament 1
averaged over six of the WHAM pointings taken around $\ell =
225\fdg5$, $b = +25\fdg5$ is displayed in Figure~\ref{fig:spectra} as
a dashed line. Because the velocity separation of the two components
($\Delta v \approx 20$ \kms) is less than the typical Doppler width of
WIM \ha\ lines (FWHM $\approx 20$ -- 30 \kms), a straightforward
two-component fit of this spectrum is not well constrained.  Instead,
we attempt to isolate the filament from gas near the LSR by
subtracting an estimate of the general WIM contribution in this
direction. The spectra of two ``background'' regions taken at the same
Galactic latitude are also displayed in Figure~\ref{fig:spectra} as
dotted lines. These directions are averages of nine pointings around
$\ell = 232\fdg0$, $b = +25\fdg5$ and 12 pointings around $\ell =
220\fdg5$, $b = +25\fdg5$. An average of these spectra is subtracted
from the average filament spectrum to produce the final filament-only
spectrum in Figure~\ref{fig:spectra} (solid line). The resulting
component is fit well by a single Gaussian with a mean of $+18\pm2$
\kms, a FWHM of $27\pm6$ \kms, and an intensity of $1.2\pm0.2$ R.

Using this method, we have determined fit parameters for several
directions along Filament 1 and 2 and present the results in
Table~\ref{tab:params}. Regions where the filaments cross are
deliberately excluded. The accuracy of the filament component
parameters depends primarily on how well the background regions
represent the true emission near $\vlsr = 0$ \kms\ in the filament
direction.  The errors of the parameters listed in
Table~\ref{tab:params} are conservatively estimated by examining
subtractions of different background directions from the filament
spectra. The fitting errors are an order of magnitude smaller and do
not contribute significantly. Note the velocity gradients in both
filaments with increasing Galactic longitude, decreasing from $\vlsr$
$+18$ \kms\ to $-25$ \kms\ in Filament 1 and increasing from $\vlsr$
$0$ \kms\ to $+18$ \kms\ in Filament 2. Also note the nearly constant
\ha\ intensities and line widths.

\section{Discussion}
\label{sec:discuss}

The intensities in Table~\ref{tab:params} are converted to emission
measures using the formula:
\begin{equation}
  \label{itoem}
  {\rm EM\ (cm}^{-6}\ {\rm pc)} = 2.75\ T_4^{0.9}\ I_{\rm H\alpha}\ 
  {\rm (R)}\ e^{2.2\,E(B-V)},
\end{equation}
where $T_4$ represents the temperature of the emitting gas in units of
$10^4$ K. With typical values for the temperature of the WIM gas, $T_4
= 0.8$ (\cite{r85}), an \ha\ intensity of 1 R corresponds to an EM of
2.25 cm$^{-6}$ pc. Since we are examining mostly high-latitude
features, we expect corrections for interstellar extinction to be
small. No such correction has been applied to the data presented
here.

To derive further physical parameters for the filaments, we must make
an assumption about the distances to them. Some evidence suggests that
the lower part of Filament 1 is at a distance of about 1 kpc, based on
its radial velocity and possible association with CMa R1/OB1. The
lower portion has a radial velocity of $+16$ \kms, which is consistent
with a kinematic distance of approximately 1 kpc at this Galactic
longitude for a value of Oort's Constant $A = 16$ \kms\ kpc$^{-1}$
(\cite{mb81}). Unfortunately, this argument is weakened by the radial
velocities of the filament's high-latitude portion, which are
``forbidden'' in the Galactic rotation model.  However, in
Figure~\ref{fig:images}d ($\vlsr = +20$ \kms), the vertical portion of
Filament 1 appears to end in the prominent \hii\ region surrounding
the star-forming association CMa R1/OB1. In fact, the center of the
brighter, arc-shaped, \ha\ emission features in this \hii\ region, as
seen on the Palomar Sky Survey red plates, is at $\ell = 225\arcdeg$,
$b = -1\arcdeg$ (\cite{ro78}). Also, the velocity centroid of the
\hii\ region (Table~\ref{tab:params}) is similar to that of the first
15\arcdeg--20\arcdeg\ of the filament away from the plane. Unless the
agreement between the spatial and velocity information of these two
regions is a coincidence, we can use the distance of the \hii\ region
as an estimate of the distance to the filament. The distance to the
CMa OB1 association has been determined photometrically by Clari\'{a}
(1974) to be $1150 \pm 140$ pc, consistent with a kinematic distance
of 800-1100 pc for the \ha\ emitting gas in the region (\cite{ro78}).
In this letter, we adopt a distance of 1 kpc for Filament 1 but
include $\dist = d/1$ kpc in the derived parameters.

At a distance of 1 kpc, the vertical extent of Filament 1 ($b \approx
+51\arcdeg$) translates to a vertical height above the Galactic
mid-plane, $Z$, of $1200\;\dist$ pc and a width ($\approx 2\arcdeg$) of
$35\;\dist$ pc. If we assume the filament is a cylinder of
uniform density gas of width $L$, then the emission measure,
\begin{equation}
  \label{em}
  {\rm EM\ (cm}^{-6}\ {\rm pc)} = \int_{0}^{L} n_e^2\,dl = n_e^2 L,
\end{equation}
of the feature can be used to estimate the density. For Filament 1,
where $L = 35\;\dist$ pc and ${\rm EM} \approx 1.1$ cm$^{-6}$ pc, $n_e
= 0.18\;\dist^{-1/2}$ cm$^{-3}$. The typical column density through
the filament is $N_e = 1.9 \times 10^{19}\;\dist^{1/2}$ cm$^{-2}$. The
mass of the material in the filament's vertical section can be
estimated by $1.4\;m_H\;Z\;L^2\;n_e = 9.2 \times 10^{3}\;\dist^{5/2}$
M$_\odot$, where the factor 1.4 is a correction for helium. If the
filament is photoionized, our observed EM implies an incident
ionizing flux of at least $\alpha_B\;{\rm EM} = 1.0 \times 10^{6}$ ph
cm$^{-2}$ s$^{-1}$. The hydrogen recombination rate within the
filament is given by $\alpha_B\;n_e^2 = 1.0 \times
10^{-14}\;\dist^{-1}$ cm$^{-3}$ s$^{-1}$, implying that the power
required to sustain this rate throughout the volume of the filament is
$(\alpha_B\;n_e^2\;Z\;L^2) \times 13.6\ {\rm eV} = 9.4 \times
10^{36}\;{\dist}^2$ erg s$^{-1}$.  The value of $\alpha_B$ used in
these calculations, $3.10 \times 10^{-13}$ cm$^3$ sec$^{-1}$, is
interpolated from Osterbrock (1989) assuming a gas temperature of 8000
K.

The narrow shape and location of Filament 1, particularly the
coincidence in location and radial velocity of one end of the filament
with an energetic source in the plane (CMa R1/OB1), suggest one
possibility for its origin---a jet-like ejection from the association.
However, given the length of the filament at the distance of CMa
R1/OB1, the apparently low velocities associated with it
(Table~\ref{tab:params}), and the near constant \ha\ intensity profile
along its entire length, it is difficult to reconcile a scenario in
which ionized hydrogen is being ejected from CMa R1/OB1 with the gas's
short recombination times. A parcel of ionized gas at these
temperatures and densities will recombine in $t_r =
(\alpha_B\;n_e)^{-1} = 1.8 \times 10^{13}\;\dist^{1/2}\ {\rm s} = 5.7
\times 10^5\;\dist^{1/2}$ yr. To reach the observed height and remain
ionized, the gas would need to be ejected at speeds in excess of
$Z/t_r = 1400\;\dist^{1/2}$ \kms.  We see no evidence for such speeds
in our data.  Furthermore, we would expect a significant gradient in
the filament's intensity as a function of height above the Galactic
plane, since gas at larger distances from the plane has had more time
to recombine.

If another source is responsible for maintaining the ionization of
Filament 1, an ejection scenario could still be plausible. If the
filament's arc shape is caused by free-fall of ejected gas back to the
Galactic plane, an initial velocity of about 70 \kms\ is required to
reach the observed height above the plane at the distance of CMa
R1/OB1. Such speeds agree better with the measured filament component
velocities in Table~\ref{tab:params}, particularly when projection
effects are considered. However, the filament's length would then
require that the ejecting source be more than $3 \times 10^7$ yr old.
This number can be compared to estimates of $3 \times 10^6$ yr for the
age of CMa OB1 by Clari\'{a} (1974) and $7.6 \times 10^5$ yr for the
age of a supernova put forth by Herbst \& Assousa (1977) as being
responsible for generating the ring-shaped \ha\ nebulosity in the
vicinity of CMa OB1/R1 and for being the progenitor of star formation
in the R1 association. In this scenario, a diffuse ionization source
is probably required to maintain a constant intensity along the
filament, since the growing distance from CMa R1/OB1 would produce an
intensity gradient if it were the sole source. The efficiency of
leaking Lyman continuum radiation from the disk is a longstanding
problem for ionizing a thicker layer of the Galaxy.  However,
Bland-Hawthorn \& Maloney (1998), Dove \& Shull (1994), and Miller \&
Cox (1993) have argued that a substantial ionizing flux can diffuse
into the Galaxy's halo from the disk. The required ionizing flux of at
least $1.0 \times 10^6$ ph cm$^{-2}$ s$^{-1}$ is consistent with these
models.

Several other phenomena may produce large, faint filaments.
Large-scale, kinematic Galactic structures such as superbubbles,
chimneys, or worms have been popular explanations for filamentary
structures seen in \ion{H}{1} as well as \ha\ (\emph{e.g.}
\cite{khr92}; \cite{ro79}). The velocity gradient of Filament 1 is
outside the range expected from a simple model of Galactic rotation
and therefore may be evidence for a dynamical influence on the
filament. However, the observed velocity gradient is inconsistent with
this feature being the edge of an expanding shell, since a shell's
projected edges should be at a constant radial velocity. Furthermore,
for the two cases presented here, we find no obvious correlation
between these \ha\ filaments and emission at other wavelengths,
including \ion{H}{1} 21 cm, Rosat All-Sky Survey X-Ray, and IRAS
bands, making it difficult to relate them to previously identified
\ion{H}{1} ``worms'' and superbubbles. Since the Leiden/Dwingeloo
\ion{H}{1} survey is sensitive to below $5 \times 10^{18}$ cm$^{-2}$
(\cite{hb97}), the ionized column density of the filament, $2 \times
10^{19}$ cm$^{-2}$, suggests that the gas is fully ionized.  Other
possibilities include a suggestion by Dupree \& Raymond (1983) that
faint, ionized trails of ionized hydrogen could be produced by
high-velocity white dwarfs.  However, the recombination time discussed
above suggests unreasonable velocities for the star.

In summary, we present the discovery of two long, faint \ha\ filaments
at high latitudes from the WHAM \ha\ survey. Their origin is not yet
identified, but the existence of such features may help to explain
processes responsible for the maintenance of the general WIM layer. As
additional portions of the \ha\ survey are reduced, new clues about
the nature of these structures may be revealed. We plan to follow
these observations with [\ion{S}{2}] and [\ion{N}{2}] observations of
this region. Additional observations will provide information about
the gas's temperature and ionization state, which are useful in
narrowing the kinds of processes that can be producing these
filaments.

We thank Kurt Jaehnig and Jeff Percival of the University of
Wisconsin's \emph{Space Astronomy Lab} for their dedicated engineering
support of WHAM; Nikki Hausen, Mark Quigley, and Brian Babler for data
reduction support; and Trudy Tilleman for essential night-sky
condition reports from Kitt Peak, which have made remote observing
possible. We acknowledge the use of NASA's \emph{SkyView} facility
(http://skyview.gsfc.nasa.gov), located at NASA Goddard Space Flight
Center and the SIMBAD database, operated at CDS, Strasbourg, France.
This work is supported by the National Science Foundation through
grants AST9619424 and AST9122701.

\newpage

\figcaption[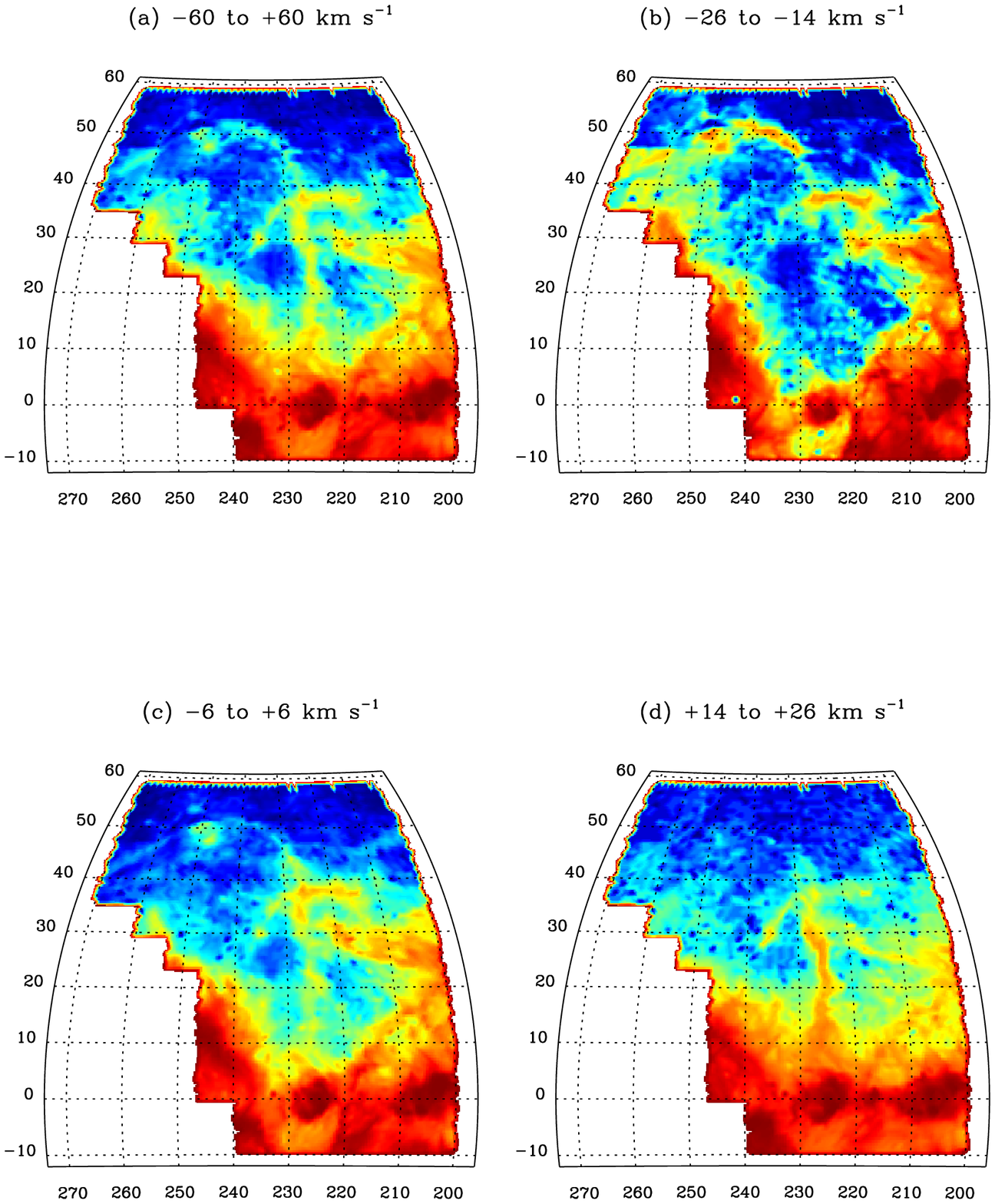] {A portion of the WHAM \ha\ Sky Survey. These
  pseudo-color images show the integrated \ha\ emission in four
  selected velocity bands.  The axes are Galactic longitude and
  latitude. (\emph{a}) $\vlsr = -60$ to $+60$ \kms.  (\emph{b}) $\vlsr
  = -26$ to $-14$ \kms.  (\emph{c}) $\vlsr = -6$ to $+6$ \kms.
  (\emph{d}) $\vlsr = +14$ to $+26$ \kms.  ``Holes'' in the images are
  pointings uncorrected for a bright stellar absorption
  line. (\emph{Preprint Note:} A color
  PostScript version of this figure can be found at
  \texttt{http://www.astro.wisc.edu/wham/papers.html})
  \label{fig:images}}

\figcaption[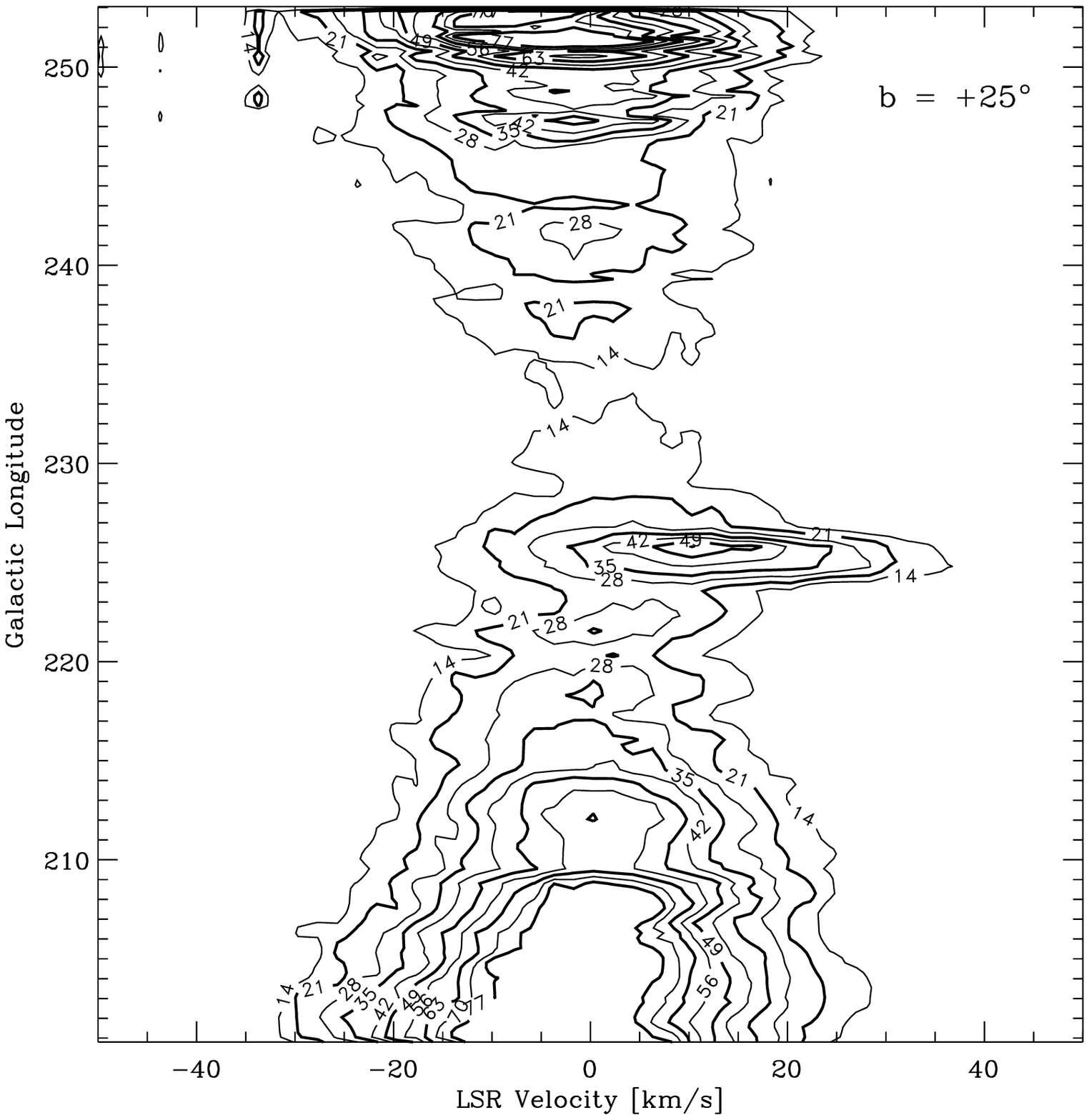] {Longitude-Velocity intensity map at $b =
  +25\arcdeg$. Contour levels are in units of $10^{-3}$ R
  (\kms)$^{-1}$. Filament 1 crosses $b = +25\arcdeg$ at $\ell =
  225\arcdeg$.
  \label{fig:lv/bcut}}

\figcaption[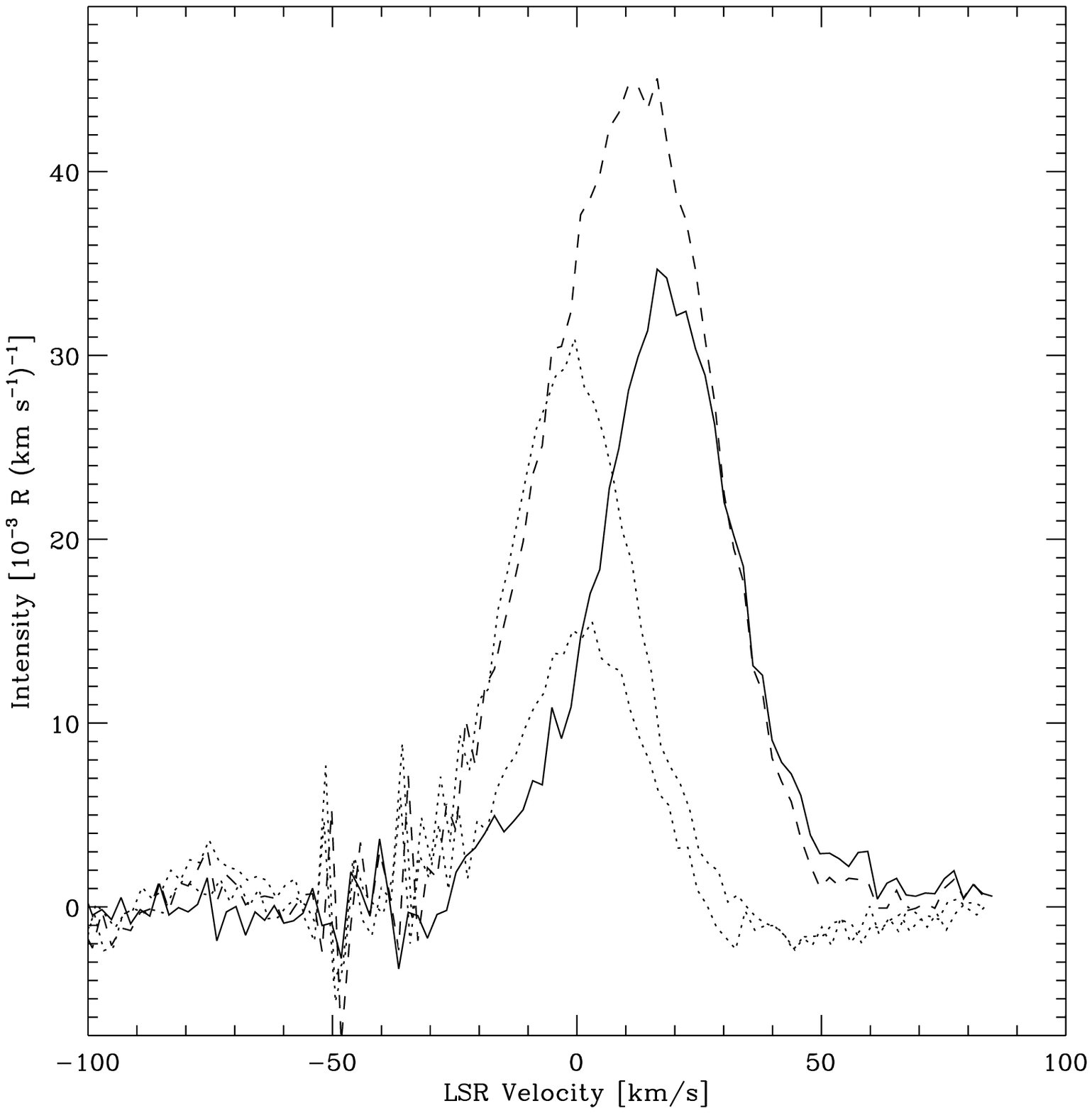] {Filament and background spectra at $b =
  +25\fdg5$. Intensity is plotted against $\vlsr$. Atmospheric lines
  have been removed. The dashed line is the original emission toward
  the filament, the dotted lines are two background directions, and
  the solid line is the background-subtracted filament emission (see
  text). The increased noise near $-40$ \kms\ is a residual of the
  geocoronal \ha\ line, whose peak emission is centered near that
  velocity.
  \label{fig:spectra}}

\newpage

\begin{deluxetable}{ccr@{$\pm$}lr@{$\pm$}lr@{$\pm$}l}
  \tablewidth{0pt}
  \tablecaption{Fitted Filament Parameters \label{tab:params}}
  \tablehead{
    \colhead{Position} &
    \colhead{\# of Avg'd} &
    \multicolumn{2}{c}{Mean} &
    \multicolumn{2}{c}{FWHM} &
    \multicolumn{2}{c}{Intensity} \nl
    \colhead{$(\ell, b)$} &
    \colhead{Spectra} &
    \multicolumn{2}{c}{(\kms)} &
    \multicolumn{2}{c}{(\kms)} &
    \multicolumn{2}{c}{(R)} \nl
    }

  \startdata

  \multicolumn{8}{c}{\underline{Filament 1}} \nl
  
  $(225\fdg5, +\phd0\fdg0)$\tablenotemark{a} & 5 & +14.0 & 0.1 &
  32.5 & 0.1 & 59.1 & 0.1 \nl

  $(227\fdg0, +10\fdg2)$ & 9 & $+18$ & 2 & 23 & 4 &
  \phd0.9 & 0.1\nl

  $(225\fdg5, +12\fdg7)$ & 9 & $+16$ & 2 & 24 & 4 &
  \phd1.1 & 0.4\nl
  
  $(225\fdg0, +15\fdg3)$ & 6 & $+18$ & 2 & 22 & 3 &
  \phd0.6 & 0.1\nl

  $(225\fdg0, +17\fdg8)$ & 6 & $+16$ & 3 & 26 & 6 &
  \phd0.6 & 0.2\nl

  $(225\fdg0, +20\fdg4)$ & 6 & $+16$ & 1 & 33 & 2 &
  \phd1.4 & 0.1\nl

  $(225\fdg5, +25\fdg5)$ & 6 & $+18$ & 2 & 27 & 6 &
  \phd1.2 & 0.2 \nl

  $(227\fdg3, +33\fdg1)$ & 12 & $+9$ & 3 & 20 & 2 &
  \phd0.6 & 0.1\nl

  $(227\fdg3, +35\fdg7)$ & 10 & $+6$ & 1 & 22 & 1 &
  \phd0.5 & 0.1\nl

  $(227\fdg3, +38\fdg2)$ & 10 & $+4$ & 3 & 25 & 2 &
  \phd0.5 & 0.4\nl

  $(227\fdg3, +40\fdg8)$ & 11 & $0$ & 4 & 24 & 6 &
  \phd0.4 & 0.2\nl

  $(227\fdg3, +43\fdg3)$ & 11 & $-1$ & 3 & 23 & 2 &
  \phd0.6 & 0.1\nl

  $(229\fdg3, +45\fdg9)$ & 9 & $-8$ & 1 & 25 & 1 &
  \phd0.4 & 0.1\nl

  $(231\fdg6, +48\fdg4)$ & 11 & $-18$ & 1 & 22 & 3 &
  \phd0.5 & 0.1\nl

  $(234\fdg0, +50\fdg5)$ & 7 & $-22$ & 1 & 22 & 4 &
  \phd0.6 & 0.1\nl

  $(239\fdg0, +51\fdg0)$ & 9 & $-25$ & 3 & 26 & 4 &
  \phd0.6 & 0.1\nl

  \multicolumn{8}{c}{\underline{Filament 2}} \nl

  $(215\fdg3, +36\fdg5)$ & 6 & $0$ & 3 & 31 & 1 &
  \phd1.2 & 0.4\nl

  $(217\fdg8, +37\fdg5)$ & 6 & $-1$ & 1 & 26 & 1 &
  \phd1.3 & 0.3\nl

  $(220\fdg5, +37\fdg5)$ & 6 & $-2$ & 1 & 27 & 2 &
  \phd1.0 & 0.4\nl

  $(230\fdg3, +35\fdg5)$ & 6 & $0$ & 4 & 22 & 6 &
  \phd0.4 & 0.2\nl
  
  $(232\fdg3, +34\fdg8)$ & 6 & $+1$ & 1 & 29 & 1 &
  \phd0.7 & 0.1\nl
  
  $(234\fdg3, +33\fdg0)$ & 6 & $+13$ & 1 & 31 & 1 &
  \phd0.4 & 0.1\nl

  $(236\fdg0, +30\fdg5)$ & 8 & $+18$ & 2 & 37 & 1 &
  \phd1.0 & 0.1\nl

  \enddata
  \tablenotetext{a}{This direction is toward the \ion{H}{2} region
    surrounding CMa R1/OB1.}
\end{deluxetable}

\newpage
 
\begin{figure}
  \begin{center}
    \leavevmode
    \plotone{fig1.ps}
  \end{center}
\end{figure}

\begin{figure}
  \begin{center}
    \leavevmode
    \plotone{fig2.ps}
  \end{center}
\end{figure}

\begin{figure}
  \begin{center}
    \leavevmode
    \plotone{fig3.ps}
  \end{center}
\end{figure}

\end{document}